\documentclass[12pt,showpacs,showkeys,amsmath,amssymb]{revtex4}
\usepackage{amsmath,amsfonts,amsthm,amscd,amssymb,latexsym}
\usepackage{bm}
\usepackage{dcolumn}
\usepackage{graphicx}
\usepackage{epstopdf}
\usepackage{color}
\usepackage{epsf}
\usepackage{epsfig}
\usepackage{graphicx, epic, eepic, color}

\newcommand{\beq}{\begin{equation}}
\newcommand{\eeq}{\end{equation}}

\newcommand\pd{\partial}

\newcommand{\negminispace}{\kern-.016667em} 

\newcommand{\half}{\kern.083333em}   
\newcommand{\quart}{\kern.0416675em}  
\newcommand{\nhalf}{\kern-.083333em}   
\newcommand{\nquart}{\kern-.0416675em}  

\newcommand{\dual}[1]{{}^\ast\nhalf\nquart#1}

\newcommand\fscalar[1]{{}^\circ{\nquart\nhalf #1}}
\newcommand\fvec[1]{{}^\dagger{\nhalf #1}} 
\newcommand\ftensor[1]{{}^\ddagger{\nquart #1}}

\newcommand\fscalarB{\fscalar{\nhalf B}}

\newcommand\fscalarE{\fscalar{\nhalf E}}

\begin{document}

\title{Twisted Gravitational Waves of Petrov Type \emph{D}}

\author{Kjell \surname{Rosquist}$^{1}$}
\email{kr@fysik.su.se}
\author{Donato \surname{Bini}$^{2,3}$}
\email{donato.bini@gmail.com}
\author{Bahram \surname{Mashhoon}$^{4,5}$}
\email{mashhoonb@missouri.edu}

\affiliation{
$^1$Department of Physics, Stockholm University, SE-106 91 Stockholm, Sweden\\
$^2$Istituto per le Applicazioni del Calcolo ``M. Picone'', CNR, I-00185 Rome, Italy\\
$^3$ICRANet, Piazza della Repubblica 10, I-65122 Pescara, Italy\\
$^4$Department of Physics and Astronomy, University of Missouri, Columbia, Missouri 65211, USA\\
$^5$School of Astronomy, Institute for Research in Fundamental
Sciences (IPM), P. O. Box 19395-5531, Tehran, Iran\\
}

\date{\today}

\begin{abstract}
Twisted gravitational waves (TGWs) are nonplanar unidirectional Ricci-flat solutions of general relativity. Thus far only TGWs of Petrov type \emph{II} are implicitly known that depend on a solution of a partial differential equation and have wave fronts with negative Gaussian curvature. A special Petrov type \emph{D} class of such solutions that depends on an arbitrary function is explicitly studied in this paper and its Killing vectors are worked out. Moreover, we concentrate on two solutions of this class, namely, the Harrison solution and a simpler solution we call the $w$-metric and determine their Penrose plane-wave limits. The corresponding transition from a nonplanar TGW to a  plane gravitational wave is elucidated. 
\end{abstract}


\maketitle

\section{Introduction}

Two recent papers~\cite{Bini:2018gbq, TGW} have discussed a class of Ricci-flat solutions of general relativity (GR) that represent nonplanar unidirectional gravitational waves with wave fronts that have negative Gaussian curvature. These \emph{twisted gravitational waves} (TGWs) are of type \emph{II} in the Petrov classification and can be represented by the spacetime interval
\begin{equation}\label{H1a}
ds^2 = -\Psi^4\,(dt^2 - dz^2 ) + \alpha^2\,\Psi^4\,\left(\frac{\partial \Psi}{\partial x}\right)^2\, dx^2 + \Psi^{-2}\, dy^2\,,
\end{equation}
where $\alpha$ is a positive constant, $u := t-z$ is the retarded null coordinate and $\Psi(u,x)$  satisfies the partial differential equation 
\beq \label{H1b}
\Psi \,\Psi_{,uu} = \upsilon(u)\,.
\eeq
Here,  $\upsilon$ is an arbitrary function of $u$ and $\Psi_{,u}:= \partial \Psi/\partial u$, etc. For $\upsilon = 0$, the general solution of Eq.~\eqref{H1b} is given by $\Psi = u\, f(x) + h(x)$, where $f(x)$ and $h(x)$ are arbitrary functions of $x$. The resulting special TGWs are of Petrov type \emph{D} and contain Harrison's  spacetime~\cite{Harrison:1959zz, DIRC}. These TGWs of type \emph{D} are the main focus of the present investigation. 

Twisted gravitational waves have nonexpanding rays and therefore belong to the Kundt class of solutions of GR~\cite{Kundt, KuTr}; for a detailed treatment of the Kundt's class, see~\cite{R1, Griffiths:2009dfa} and the references cited therein. It is therefore possible to put metric~\eqref{H1a} in the standard Kundt form; see, Appendix A of Ref.~\cite{TGW}. It is likely that TGWs are known in certain other forms and expressed in other coordinate systems. In particular, Kinnersley has employed the Newman-Penrose method to characterize all Ricci-flat solutions of Petrov type \emph{D}~\cite{Kin}. 

We assume throughout that the waves propagate in the $z$ direction. In $(t, z, x, y)$ coordinates, simple twisted gravitational waves of type \emph{D} have a metric of the form 
\begin{equation}\label{H1}
 ds^2 = (u\,f+h)^4 (- dt^2 + dz^2) + \alpha^2\,(u\,f + h)^4 \left(u\, \frac{df}{dx} + \frac{dh}{dx}\right)^2 \,dx^2 + (u\,f + h)^{-2}\,dy^2\,,
\end{equation}
which follows from metric~\eqref{H1a} with $\Psi(u, x) = u\,f + h$. It is evident from this form of the metric that passing to a coordinate system $(u,v,x,y)$, $\partial_v$ and $\partial_y$ are Killing vector fields, where $v := t+z$ is the advanced null coordinate. 

Let us first assume that $f$ is not a constant and use the freedom in the choice of the $x$ coordinate to define $f(x) := \mathcal{X}$ as the new $x$ coordinate and $h(x) := \mathcal{X}\,\mathcal{H}(\mathcal{X})$; then, metric~\eqref{H1} takes the form
\begin{equation}\label{H2}
 ds^2 = \mathcal{X}^4\,(u+\mathcal{H})^4 (- dt^2 + dz^2) + \alpha^2\,\mathcal{X}^4\,(u + \mathcal{H})^4\, \left(u + \mathcal{H} + \mathcal{X}\, \frac{d\mathcal{H}}{d\mathcal{X}}\right)^2 \,d\mathcal{X}^2 + \frac{1}{\mathcal{X}^2\,(u + \mathcal{H})^2}\,dy^2\,,
\end{equation}
where $\mathcal{H}$ is an arbitrary function of $\mathcal{X}$. Next, we define $\mathcal{X}^3 := X$ in order to get the final form for the metric, namely, 
\begin{equation}\label{H3} 
 ds^2 = X^{4/3}\,(u+H)^4 (- dt^2 + dz^2) + \alpha_0^2\,(u+ H)^4 \left(u + H  +3\,X\, \frac{dH}{dX}\right)^2 \,dX^2 + \frac{1}{X^{2/3}\,(u + H)^2}\,dy^2\,,
\end{equation}
where $H(X) := \mathcal{H}(\mathcal{X})$ and $\alpha_0 = \alpha/3$. 

Harrison's TGW corresponds to the particular case of a constant magnitude for $H$. That is, for constant $H$ we define $U = T-Z := (u+H)^5 / (5\,C_0)$, where $C_0 \ne 0$ is a constant; then, defining a new advanced null coordinate $V = T+Z$ via $C_0\, v := V$, a new $Y$ coordinate via $y/(5\,C_0)^{2/5} := Y$ and setting $\alpha_0^2\,(5\,C_0)^{6/5} = 1$, we get Harrison's TGW metric~\cite{Harrison:1959zz}
\begin{equation}\label{H4} 
 ds^2 = X^{4/3}\, (- dT^2 + dZ^2) + U^{6/5}\,dX^2 + \frac{1}{X^{2/3}\,U^{2/5}}\,dY^2\,.
\end{equation}
Thus for $H \ne$  constant, we have in Eq.~\eqref{H3} a generalization of Harrison's TGW spacetime that depends on an arbitrary function $H(X)$. 

Let us next assume that $f$ is in fact a nonzero constant, namely,  $f = \chi \ne 0$. Then, the freedom in the choice of the $x$ coordinate implies, as before, that we can set $h(x) = \eta\, x$, where $\eta >0$ is a constant such that $\alpha^2 \,\eta^2 = 1$. The TGW metric~\eqref{H1a} in this case has the simple form
\begin{equation}\label{H5}
 ds^2 = W^4 (-dt^2 + dz^2) + W^4\, dx^2 + W^{-2} \,dy^2\,,
\end{equation}
where $W = \chi \,u + \eta\,x$. If we now replace $x$ by $x+1/ \eta$, $W$ becomes $1+W$; that is, with a simple translation in $x$, say, $W$ can be changed into 
$W = 1+ \chi \,u + \eta\,x$, which is a TGW on the Minkowski spacetime background. For $\chi = 0$, metric~\eqref{H5} reduces to a \emph{static} Kasner solution~\cite{Bini:2018gbq, TGW}.

Background material in connection with exact gravitational waves in GR can be found in Refs.~\cite{R1, Griffiths:2009dfa} and the references cited therein. As mentioned before, the simple type \emph{D} solutions representing TGWs under consideration here are all actually known in other contexts. They are in Kundt's class of nondiverging solutions discussed in Section 31.5.2 of Ref.~\cite{R1} and Section 18.6 of Ref.~\cite{Griffiths:2009dfa}. According to Section 31.5.2 of Ref.~\cite{R1}, all such vacuum solutions have four Killing vectors with a corresponding timelike three-dimensional orbit. The general form of the metric under consideration in this paper is given by Eq.~\eqref{H1}, which clearly admits $\partial_v$ and $\partial_y$ as Killing vector fields. We find the other two Killing vectors by first studying the spacetimes associated with Harrison's TGW and the $w$-metric, which is a special case of Eq.~\eqref{H5} with $\chi = \eta = 1$.

\section{Harrison's TGW}

Inspection of Harrison's solution, which we henceforth write in the form
\begin{equation}\label{H6} 
 ds^2 = x^{4/3}\, (- dt^2 + dz^2) + u^{6/5}\,dx^2 + \frac{1}{x^{2/3}\,u^{2/5}}\,dy^2\,,
\end{equation}
reveals simply that it has null ($\partial_v= \partial_t+\partial_z$) and spacelike ($\partial_y$) Killing vector fields that are hypersurface-orthogonal as well as a homothetic vector field $5\, t\, \partial_t + 6\,x\, \partial_x\, + 12\, y\, \partial_y+ 5\,z\, \partial_z$.

\subsection{Principal Null Directions}

Plane gravitational waves are of type \emph{N} in the Petrov classification and have parallel rays such that the four principal null directions of the Weyl tensor coincide. The resulting principal null direction is parallel to the null propagation vector of the wave and is thus normal to the wave front. With respect to the set of fiducial observers that are all at rest in space, timelike geodesics in such spacetimes generally line up at late times in the direction of wave propagation with Lorentz factors that asymptotically tend to infinity. This \emph{cosmic jet} feature of plane gravitational waves has been the subject of previous investigations~\cite{Bini:2014esa, Bini:2017qnd, Tucker:2016wvt, Tucker:2018xle}. 

\emph{Simple} TGWs are of type \emph{D} in the Petrov classification and thus have two repeated principal null directions. For the Harrison spacetime, these are given by
\begin{eqnarray}\label{H7}
\bar{n}&=&x^{-2/3}(\partial_t+\partial_z)\,,\nonumber\\
\bar{l} &=&\frac{1}{2}\left[\frac{9}{25 u^{4/5}} +\frac{1}{ x^{2/3}} \right]\partial_t  -\frac{3x^{1/3}}{5u} \partial_x  +\frac{1}{2}\left[\frac{9}{25 u^{4/5}}-\frac{1}{x^{2/3}} \right]\partial_z\,,
\end{eqnarray}
which satisfy the condition $\bar{n}^\mu\,\bar{l}_\mu = -1$ and can be considered to be the two real legs of a null tetrad frame. We note that $\bar{n}$ is parallel to the direction of wave propagation and hence normal to the wave front, while $\bar{l}$ is in some oblique direction with respect to the direction of wave propagation. Let us now consider geodesics of Harrison's gravitational field. Investigations of the asymptotic behavior of timelike geodesics with respect to fiducial observers that are all at rest in space indicate that the cosmic-jet property persists in such simple TGWs that propagate along the $z$ direction, but the cosmic jet propagates in the $(z,x)$ plane along a direction  that deviates from the direction of 
wave propagation~\cite{Bini:2018gbq, Bini:2017qnd}. Thus this TGW of Petrov type \emph{D} still exhibits a single characteristic cosmic jet, but it propagates obliquely with respect to the direction of wave propagation.

\subsection{Two More Killing Vectors}

In the Harrison metric~\eqref{H6}, let us consider the coordinate transformation $(t, z, x, y) \mapsto (v', w', x', y')$, where
\begin{equation}\label{H8}
   v' = 3^{4/3}\, (t-z)^{-1/5}\,,\qquad  w' = 5\,(t+z)\,, \qquad
   x' = 3^{5/3}\, x^{1/3}\,, \qquad y' = 3^{1/3}\, y\,.
\end{equation}
Then, Harrison's metric takes the form
\begin{equation}\label{H9}
 ds^2 = \frac{x'^4}{v'^6}\, (d v' d w' + d x'^2) + \frac{v'^2}{x'^2}\, d y'^2\,.
\end{equation}
In terms of the new coordinates, the two known Killing vectors are $\kappa = \partial_{w'}$ and $\varsigma = \partial_{y'}$.  Furthermore, the $y' = \text{constant}$ hypersurfaces are manifestly conformally flat. 
 
The new form of Harrison's metric~\eqref{H9} is invariant under the scaling symmetry $(v', w', x', y')\mapsto (\beta\,v', \beta\, w', \beta \,x',  y')$, where $\beta \ne 0$ is a constant. This symmetry  corresponds to the existence of a third Killing field $n$, 
\begin{equation}\label{H10}
   n = v'\, \pd_{v'} + w'\, \pd_{w'} + x'\, \pd_{x'}\,,
\end{equation}
which has the squared norm $n_\alpha\, n^{\alpha} = v'^{-6} x'^4 (v'\,w'+x'^2)$.

All metrics in Kundt's class admit a $G_4$ symmetry group acting on a 3-dimensional timelike hypersurface, see Section 31.5.2 of Ref.~\cite{R1}. Therefore, we search for a fourth Killing vector field $m$ of the form
\begin{equation}\label{H11}
  m = P (v', w', x', y') \,\kappa + Q(v', w', x', y')\, n\,,
\end{equation}
where $P$ and $Q$ should be determined from the Killing equation 
\begin{equation}\label{H12}
  m_{\alpha, \beta} + m_{\beta, \alpha} = 2\,\Gamma_{\alpha \beta}^{\mu}\,m_{\mu}\,.
\end{equation}
That is, 
\beq\label{H12a}
m_\mu = P\,\kappa_\mu + Q\,n_\mu\,,
\eeq
where $\kappa_\mu$ and $n_\mu$ are given by
\beq\label{H12b}
\kappa_\mu\,dx^\mu = \frac{x'^4}{2v'^6}\,dv'\,, \qquad n_\mu\,dx^\mu = \frac{x'^4}{2v'^6}\,d(v'w' + x'^2)\,.
\eeq
Substituting Eq.~\eqref{H12a} in the Killing Eq.~\eqref{H12} and noting that $\kappa$ and $n$ satisfy the Killing equation as well, we find
\beq\label{H12c}
P_{,\mu}\,\kappa_\nu + P_{,\nu}\,\kappa_\mu + Q_{,\mu}\,n_\nu + Q_{,\nu}\,n_\mu = 0\,.
\eeq
Writing out the ten components of this equation, we find that most of the equations simply imply that $P$ is independent of $y'$ and $Q$ only depends on $v'$; furthermore, the three components $(\mu, \nu) = (v', v')$, $(v', w')$ and $(v', x')$ can then be written as 
\beq \label{H12d}
\frac{\partial P}{\partial v'} + w'\,\frac{dQ}{dv'} = 0\,, \qquad \frac{\partial P}{\partial w'} + v'\,\frac{dQ}{dv'} = 0\,,\qquad \frac{\partial P}{\partial x'} + 2\,x'\,\frac{dQ}{dv'} = 0\,,
\eeq
respectively. The integrability condition for the first two of these three equations implies
\beq\label{H12e}
Q-Q_1 = v'\,\frac{dQ}{dv'}\,
\eeq
or $Q = Q_0\,v' +Q_1$, where $Q_0$ and $Q_1$ are integration constants.  It is then straightforward to solve the three equations and conclude that $P = - Q_0\,(v'w' + x'^2) + P_0$, where $P_0$ is a new integration constant.  It follows from these results that the fourth Killing vector field is given,  up to a constant multiplicative factor, by
\begin{equation}\label{H13}
 m = -  (v'\,w' + x'^2)\,\kappa + v'\, n\,.
\end{equation} 
Hence, the fourth Killing vector is \emph{null} and can be expressed as
\begin{equation}\label{H14}
   m = v'^2\, \pd_{v'} - x'^2\, \pd_{w'} + v'x' \,\pd_{x'}\,. 
\end{equation}
The corresponding Killing 1-forms for the Harrison solution are thus
\begin{equation}\label{H15}
 \begin{split}
   \tilde \kappa &= \frac{x'^4}{2v'^6} \,d v' \,, \\[5pt]
   \tilde m &= \frac{x'^4}{2v'^6}\,(-x'^2 d v' + 2\,v\,'x' \,d x' + v'^2 d w')
             = \frac{x'^4}{2v'^4}\, d \Bigl(\frac{v'\,w' + x'^2}{v'} \Bigr)\,, \\[5pt]
   \tilde n &= \frac{x'^4}{2v'^6} \,(w' d v' + v' d w' + 2x' d x')
             = \frac{x'^4}{2v'^6}\, d(v'\,w' + x'^2)\,, \\[5pt]
   \tilde \varsigma &= \frac{v'^2}{x'^2} \, d y'\,.
 \end{split}
\end{equation}

It is interesting to note that in terms of the original $(t, z, x, y)$ coordinates, $\kappa = 10^{-1} (\partial_t + \partial_z)$ and $\varsigma = 3^{-1/3}\,\partial_y$, while the new Killing vectors are 
\begin{equation}\label{H16}
   n = -(2t-3z)\, \pd_{t} + 3x\, \pd_{x} + (3t-2z)\, \pd_{z}\,, \qquad  m = -\frac{3^{4/3}}{10}\, \mathbb{K}\,,
\end{equation} 
where $\mathbb{K}$ is given by 
\beq\label{H17a}
\mathbb{K}= [9\, x^{2/3} + 25\, (t-z)^{4/5}]\,\partial_t - \frac{30\, x}{(t-z)^{1/5}}\,\partial_x +[ 9\, x^{2/3} - 25\, (t-z)^{4/5}]\,\partial_z\,.
\eeq
Moreover, a 2-dimensional subspace invariant with respect to all of the Killing vectors is defined by $(dF = 0, dy = 0)$, where $F(t, z, x, y) = (t-z)^{3/5}\,x$. Therefore, the gradient of $F$ is perpendicular to all the Killing vectors. 

The two repeated principal null directions of Harrison's spacetime are parallel to the two null Killing vector fields; that is, 
\beq \label{H17b}
\bar{n} = 10\sqrt{2}\,x^{-2/3}\,k\,, \qquad \bar{l} = -\frac{10^{-1}\sqrt{2}}{3^{4/3}}\,x^{-2/3}\,u^{-4/5}\,m\,. 
\eeq
Of the other two Killing vector fields, one is spacelike ($\partial_y$), while $n$ could be timelike, null or spacelike, since
\beq \label{H17c}
n_\alpha\,n^\alpha = 5\,(t^2-z^2)\,x^{4/3} + 9\,(t-z)^{6/5}\,x^2\,.
\eeq

The four-parameter group of isometries of Harrison's spacetime contains a three-parameter subgroup characterized by the Killing vectors $\kappa$, $m$ and $n$, which are all perpendicular to the $y$ direction. For $(\kappa, m, n)$, the commutation relations are 
\begin{equation}\label{H18}
[\kappa,m] = 0\,, \qquad [\kappa,n] = \kappa\,, \qquad [m,n] = -m\,.
\end{equation}
This is the Lie algebra of the Lorentz group of 2-dimensional Minkowski spacetime. As a 3-dimensional group it can also be classified as being of Bianchi type $VI_0$. We note that the orbits of these three Killing symmetries form a collection of timelike 2-dimensional surfaces of (multiple) transitivity (\emph{cf}. Ref.~\cite{Shikin}).

The timelike geodesics of Harrison's spacetime have been studied in Section VI of Ref.~\cite{Bini:2017qnd}. The projection of the 4-velocity of a timelike (or null) geodesic on the new Killing vectors $n$ and $m$ should result in constants of the motion $\mathcal{C}_n$ and $\mathcal{C}_m$, respectively. A detailed investigation, based on the results of Ref.~\cite{Bini:2017qnd},  reveals that $\mathcal{C}_n$ is indeed a constant, while 
\beq\label{H19}
\mathcal{C}_m = \frac{3^{10/3}}{10}\, \eta_0\,E\,,
\eeq
where $\eta_0$ and $E$ are constants defined in Eqs. (87) and~(95) of Ref.~\cite{Bini:2017qnd}.

\section{$w$-Metric}

A special case of metric~\eqref{H5} for $\chi = \eta = 1$ is the $w$-metric, namely, 
\begin{equation}\label{Q1}
 ds^2 = w^4 (-dt^2 + dz^2) + w^4\, dx^2 + w^{-2} \,dy^2\,, \qquad w = u+x\,.
\end{equation}
It follows from simple inspection that the spacetime associated with the $w$-metric has one null and two independent spacelike Killing vector fields that are all hypersurface-orthogonal and are given by
\begin{equation}\label{Q2}
k =  \partial_v =  \partial_t +  \partial_z\,, \qquad p = \partial_x+  \partial_z\,, \qquad  \sigma =\partial_y\,,
\end{equation}
respectively, as well as a homothetic vector field
\begin{equation}\label{Q3}
 t\, \partial_t + x\, \partial_x\, + 4\, y\, \partial_y+ z\, \partial_z\,.
\end{equation}

An alternative interpretation of the $w$-metric involves writing the metric in the form 
\begin{equation}\label{Q4}
 ds^2 = w^4 (-dt^2 + dx^2) + w^{-2} \,dy^2 + w^4\, dz^2 \,, \qquad w = t+x-z\,,
\end{equation}
which is a twisted gravitational wave propagating in the $-x$ direction. The metric at the wave front, i.e. $t+ x$ = constant, has essentially the same Gaussian curvature as the standard form of the $w$-metric.  The normal to the wave front is the null Killing propagation vector $k-p = \partial_t - \partial_x$. Both interpretations are equally valid and provide clear justification for the fact that the \emph{cosmic jet} in this case propagates obliquely in the $(z,x)$ plane~\cite{Bini:2018gbq}. Let us briefly recall here that in a plane-wave spacetime, the motion of timelike geodesics with respect to a set of fiducial observers has the cosmic-jet property, namely, free test particles generally line up in the direction of wave propagation with Lorentz factors that asymptotically approach infinity. In TGWs of type \emph{D}, however, the direction of the cosmic jet deviates from the direction of wave propagation. Thus in the present case the cosmic jet direction deviates from both the $z$ and $x$ directions, as expected.

\subsection{Principal Null Directions}

The two repeated principal null directions of the $w$-metric are
\begin{equation}\label{Q4}
\bar{n} =\frac{1}{w^2}(\partial_t +\partial_z)\,, \qquad \bar{l} = \frac{1}{w^2}(\partial_t-\partial_x)\,,
\end{equation}
where, as before, $\bar{n}^\mu\,\bar{l}_\mu = -1$, and hence $\bar{n}$ and $\bar{l}$ can be considered to be the two real legs of a null tetrad frame. As in the case of Harrison's spacetime, these principal null directions are parallel to the two null Killing vectors of the $w$-metric; that is, 
\begin{equation}\label{Q5}
 \bar{n} = \frac{1}{w^2}\, k\,, \qquad \bar{l} =  \frac{1}{w^2}\,(k-p)\,.
\end{equation}

\subsection{Fourth Killing Vector}

As in the case of Harrison's spacetime, we assume that the fourth Killing vector field $q$ is a linear combination of $k = \partial_t + \partial_z$ and $p = \partial_x + \partial_z$. Hence, 
\beq\label{Q6}
q = A(t, z, x, y)\, k + B(t, z, x, y)\,p\,,
\eeq
where $A$ and $B$ should be determined from Killing's equation. Both $k$ and $p$ satisfy the Killing equation; therefore, we have
\beq\label{Q7}
A_{,\mu}\, k_\nu + A_{,\nu}\, k_\mu + B_{,\mu}\, p_\nu + B_{,\nu}\, p_\mu = 0\,,
\eeq
where $k_\mu$ and $p_\mu$ are given in $(t, z, x, y)$ coordinates by
\beq\label{Q8}
k_\mu = w^4\,(- 1, 1, 0, 0)\,, \qquad p_\mu = w^4\,(0, 1, 0, 1)\,.
\eeq
Most of the ten equations contained in Eq.~\eqref{Q7} simply imply that $A$ is only a function of $x$ and $z$, while $B$ is only a function of $t$ and $z$. The remaining equations may be written as
\beq \label{Q9}
\frac{\partial A}{\partial x} = \frac{\partial B}{\partial t} = \frac{\partial A}{\partial z} = -\frac{\partial B}{\partial z}\,.
\eeq
These results simply imply that $A = a (x) + e(z)$ and $B = b(t) - e(z)$, where $a$, $b$ and $e$ are all linear functions of their arguments with the same constant derivative. Thus the fourth independent Killing vector field is given by $q=A\,k + B\,p$, where
\beq \label{Q10}
A = x + z\,, \qquad B = t - z\,.
\eeq
It follows that 
\beq \label{Q11}
q = (x+z)\,\partial_t + (t-z)\,\partial_x + (t+x)\,\partial_z\,
\eeq
and its square norm is
\beq \label{Q12}
q_\alpha\,q^\alpha = 2\,w^4 (t-z)(t+x)\,,
\eeq
so that $q$ could be timelike, null or spacelike.

The four-parameter group of isometries of the $w$-metric contains a three-parameter subgroup characterized by the Killing vectors $k$, $p$ and $q$, which are all perpendicular to the $y$ direction.
For $(k, p, q)$, the commutation relations are 
\begin{equation}\label{Q13}
   [k,p] = 0\,, \qquad [k,q] = k\,, \qquad [p,q] = 2\,k- p\,.
\end{equation}
This Lie algebra corresponds to the Lorentz group of 2-dimensional Minkowski spacetime, just as in the case of Harrison's TGW. As in that case, the corresponding symmetry group acts transitively on 2-dimensional timelike subspaces.  
In fact, Eq.~\eqref{Q13} is related to Eq.~\eqref{H18} in the Harrison case via $k = \kappa$,  $p = \kappa + \gamma_1\,m$ and $q = n + \gamma_2\,\kappa + \gamma_3\,m$, where $\gamma_1 \ne 0$, $\gamma_2$ and $\gamma_3$ are constants. 

The projection of the 4-velocity of a timelike (or null) geodesic on the fourth Killing vector $q$ should produce a new constant of the motion $\mathcal{C}_q$. Indeed, $q$ generates a constant of the motion for timelike and null geodesics of the $w$-metric given by
\beq \label{Q14}
\mathcal{C}_q = C_0\,(t-z) -C_v\,(x+z)\,,
\eeq
where $C_0$ and $C_v$, defined in Eq. (69)  of Ref.~\cite{Bini:2018gbq}, are constants of the motion due to the existence of  the Killing vector fields $p$ and $k$, respectively.  In fact, $\mathcal{C}_q$ is a constant of the motion for $\lambda_0 = \lambda = 1$ in Section IV of Ref.~\cite{Bini:2018gbq}.

\subsection{Static Representation of the $w$-Metric}

Let us start with the following form of the $w$-metric
\beq \label{W1}
ds^2 = -w^4 du\,dv + w^4\, dx^2 + w^{-2} \,dy^2\,, \qquad u := t-z\,, \qquad v :=t+z\,, \qquad w = t-z+x\,.
\eeq
Consider the coordinate transformation
\beq \label{W2}
(u, v, x, y) \mapsto (\tau, \eta, w, y)\,, \qquad u = -\tau + \eta\,, \qquad v = -2(\tau + w)\,, \qquad x = \tau - \eta +w\,.
\eeq
The $w$-metric then takes on the static form
\begin{equation}\label{W3}
ds^2 = w^4\,(- d\tau^2 + d\eta^2 + dw^2) + w^{-2} dy^2\,. 
\end{equation}
In these coordinates, the Killing vectors are
\begin{equation}\label{W4}
   \pd_\tau \ ,\quad \pd_\eta \ ,\quad 
   \eta\half \pd_\tau + \tau \quart \pd_\eta \ ,\quad \pd_y\,. 
\end{equation}
The first three of these correspond precisely to the Killing vectors of a 2-dimensional Minkowski spacetime expressed in standard coordinates.

Let us now consider the coordinate transformation
\beq \label{W5}
(\tau, \eta, w, y) \mapsto (T, Z, X, Y)\,, \quad 3^{4/3}\,\tau = T\,, \quad 3^{4/3}\, \eta = Z\,, \quad \frac{w^3}{3} = X\,, \quad 3^{-2/3}\,y = Y\,.
\eeq
Then, metric~\eqref{W3} reduces to the static type \emph{D} Kasner metric
\begin{equation}\label{W6}
ds^2 = -X^{4/3} (dT^2-dZ^2) + dX^2  + X^{-2/3} dY^2\,. 
\end{equation}
The general form of the spacelike Kasner metric  is~\cite{Bini:2018gbq}
 \begin{equation}\label{W7}
 ds^2 = - x^{2p_1} dt^2 +dx^2 + x^{2p_2} dy^2 + x^{2p_3} dz^2\,,
\end{equation}
where $p_1+p_2+p_3=p_1^2+p_2^2+p_3^2 =1$. In our case, the $w$-metric is equivalent to the static Kasner metric with 
\begin{equation}\label{W8}
 p_1=p_3 =\frac{2}{3}\,, \qquad p_2 = - \frac{1}{3}\,.
\end{equation}

\section{TGWs of Type \emph{D}}

The spacetimes that we have investigated thus far, namely, those corresponding to Harrison's TGW and the $w$-metric, each contain four Killing vector fields of which two are null and parallel to the repeated principal null directions of the corresponding Weyl tensor. This leads us to conjecture that this could be a characteristic feature  of all TGWs of Petrov type \emph{D} under investigation in this paper. Let us therefore consider the general form of the metric given by Eq.~\eqref{H1} and write it in $(u, v, x, y)$ coordinates as
\begin{equation}\label{Z1}
   ds^2 = - \Psi^4 du \,dv + \alpha^2\,\Psi^4\,\left(\frac{\partial \Psi}{\partial x}\right)^2\, dx^2 + \Psi^{-2}\, dy^2\,,
\end{equation}
where $\Psi(u,x)= u\,f(x) + h(x)$. Using the standard algorithm, we find that the two principal null directions are given in this case by
\begin{equation}\label{Z2}
   \bar{n} = \Psi^{-2}\,\pd_v \,, \qquad \bar{l} = 2\,\Psi^{-2}\,\left[\pd_u + \alpha^2 f^2(x) \,\pd_v 
       -\frac{f(x)}{u f'(x) + h'(x)} \,\pd_x\right]\,,
\end{equation}
where $\bar{n}^\mu\,\bar{l}_\mu = -1$, $f'(x) := df(x)/dx$ and $h'(x) := dh(x)/dx$.  We can compare Eq.~\eqref{Z2} with Eq.~\eqref{H7} for the Harrison TGW spacetime. It turns out, in agreement with our conjecture, that $\Psi^2\,\bar{l}$ is indeed a Killing vector field in the spacetime given by metric~\eqref{Z1}. 

The null vectors $k = \pd_v$ and $M := \Psi^2\,\bar{l}/2$,
\begin{equation}\label{Z2a}
 M =\pd_u + \alpha^2 f^2(x) \,\pd_v -\frac{f(x)}{u f'(x) + h'(x)} \,\pd_x\,,
\end{equation}
are geodesic Killing vector fields in the general spacetime under consideration, while $\sigma = \pd_y$ is a spacelike Killing vector field. As in Sections II and III, we now assume that the fourth Killing vector field $N$ is in the timelike plane spanned by $k$ and $M$; hence, we can write
\begin{equation}\label{Z3}
 N_\mu = \mathcal{P}(u,v,x,y)\,k_\mu - \mathcal{Q}(u,v,x,y)\,M_\mu\,.
\end{equation}
Here, $\mathcal{P}$ and $\mathcal{Q}$ can be determined from Killing's equation, namely, 
\beq\label{Z4}
\mathcal{P}_{,\mu}\,k_\nu + \mathcal{P}_{,\nu}\,k_\mu - \mathcal{Q}_{,\mu}\,M_\nu - \mathcal{Q}_{,\nu}\,M_\mu = 0\,,
\eeq
see Eqs.~\eqref{H11}--\eqref{H12c}. Of the ten relations in Eq.~\eqref{Z4}, the first four involving $(\mu , \nu) = (u,u), (u, v), (u,x)$ and $(u, y)$ imply
\begin{eqnarray}\label{Z5}
\frac{\partial \mathcal{P}}{\partial u}& - &\alpha^2 f^2\,\frac{\partial \mathcal{Q}}{\partial u} = 0\,,\nonumber \\
\frac{\partial \mathcal{P}}{\partial v}& - &\frac{\partial \mathcal{Q}}{\partial u} - \alpha^2 f^2\,\frac{\partial \mathcal{Q}}{\partial v} = 0\,,\nonumber \\
\frac{\partial \mathcal{P}}{\partial x}& - &2\,\alpha^2 f\,(u f' + h')\,\frac{\partial \mathcal{Q}}{\partial u} - \alpha^2 f^2\,\frac{\partial \mathcal{Q}}{\partial v} = 0\,,\nonumber \\
\frac{\partial \mathcal{P}}{\partial y}& - &\alpha^2 f^2\,\frac{\partial \mathcal{Q}}{\partial y} = 0\,,
\end{eqnarray}
respectively. On the other hand, the other six relations in Eq.~\eqref{Z4} simply imply that  $\mathcal{Q}$ is only a function of $u$. It is then straightforward to see from Eq.~\eqref{Z5}  that $\mathcal{P}$ is independent of $y$ and is given by
\beq\label{Z6}
\mathcal{P} = \frac{d\mathcal{Q}}{du}\,\left[\alpha^2 f^2(x)\,u + v+ 2\,\alpha^2\int^x  \!\!\!f(\xi)h'(\xi) d\xi\right]\,.
\eeq
Thus the fourth Killing vector is of the form 
\beq\label{Z7}
N = \frac{d\mathcal{Q}}{du}\,\left[\alpha^2 f^2(x)\,u + v+ 2\,\alpha^2\int^x  \!\!\!f(\xi)h'(\xi) d\xi\right]\,k -\mathcal{Q}(u)\,M\,
\eeq
and depends on an arbitrary function $\mathcal{Q}(u)$. If $\mathcal{Q}(u)$ is a constant, then we get back the third Killing vector; otherwise, we note that in  metric~\eqref{Z1}, we can simply replace $u$ by $\mathcal{Q}(u)$ and the form of the metric remains invariant. It follows that we can set $\mathcal{Q}(u) = u$ with no loss in generality. The end result is that the three Killing vectors $k$, $M$ and 
\beq\label{Z8}
N = \left[\alpha^2 f^2(x)\,u + v+ 2\,\alpha^2\int^x  \!\!\!f(\xi)h'(\xi) d\xi\right]\,k-u\,M
\eeq
are all hypersurface-orthogonal and form a three-parameter subgroup perpendicular to the $y$ direction. That is, for $(k, M, N)$ the commutation relations are 
\begin{equation}\label{Z9}
[k,M] = 0\,, \qquad [k,N] = k\,, \qquad [M,N] = -M\,,
\end{equation}
just as in the cases of Harrison's TGW spacetime and the $w$-metric. Therefore, as before, this Lie algebra corresponds to the Lorentz group of 2-dimensional Minkowski spacetime. This symmetry subgroup therefore acts transitively on timelike 2-dimensional subspaces also in this general case of TGW spacetime of type \emph{D}. It should be noted that the isomorphic Lie algebras~\eqref{H18},~\eqref{Q13} and~\eqref{Z9} can all be transformed by a linear transformation with constant coefficients to coincide with the Lie algebra of 2-dimensional Minkowski spacetime in the standard coordinates given in Eq.~\eqref{W4}.

\section{Penrose Limit}

According to Penrose, near any null geodesic in a general relativistic spacetime, the spacetime metric takes the form of a plane gravitational wave in the Penrose limit~\cite{Penrose}. The Penrose limit has been discussed by a number of authors, see Refs.~\cite{BFS, Perlick} and the references cited therein. 

It is interesting to investigate the Penrose limit in the case of twisted gravitational wave spacetimes. In this limit, the negative curvature of the wave front turns to zero and the nonplanar wave front of the TGW becomes plane.

\subsection{Harrison's TGW}

Let us first consider Harrison's TGW and write metric~\eqref{H6} in the form 
\beq \label{N1}
ds^2 = -x^{4/3}\, du\,dv + u^{6/5}\, dx^2 + x^{-2/3}\,u^{-2/5} \,dy^2\,, \qquad u := t-z\,, \qquad v :=t+z\,,
\eeq
where we assume henceforth that $0 \le x <\infty$ and $-\infty < u < \infty$. We recall that a spacetime singularity occurs at either $x = 0$ or $u = 0$.
The null geodesic about which we take the Penrose limit is 
\beq \label{N2}
k = \partial_v\,.
\eeq
To render metric~\eqref{N1} in a form appropriate for the Penrose procedure~\cite{Penrose}, we consider the coordinate transformation 
\beq \label{N3}
(u, v, x, y) \mapsto (u', v, \xi, y)\,, \qquad u= x^{-4/3}\,u'\,,\qquad x = e^\xi\,,
\eeq
which turns Eq.~\eqref{N1} into
\beq \label{N4}
ds^2 = - du'\,dv + \frac{4}{3}\,u'\,dv\,d\xi + u'\,^{6/5}\,e^{2\xi/5}\, d\xi^2 + u'\,^{-2/5}\,e^{-2\xi/15} \,dy^2\,.
\eeq
It is now possible to implement the procedure suggested by Penrose~\cite{Penrose}. That is, we define new coordinates $(U,V,X,Y)$ such that 
\beq \label{N5}
U = u'\,,\qquad \Omega^2\,V = v\,,\qquad \Omega\,X = \xi\,,\qquad \Omega\,Y = y\,,
\eeq
where $\Omega$ is a positive constant. Then metric~\eqref{N4} takes the form
\beq \label{N6}
ds^2 = \Omega^2\,\left[-dU\,dV + \frac{4}{3}\,\Omega\,U\,dV\,dX + U^{6/5}\,e^{2\Omega X/5}\,dX^2 +U^{-2/5}\,\,e^{-2\Omega X/15}\,dY^2\right]\,.
\eeq
Let us define the conformally related metric $d\bar{s}^2$ such that 
\beq \label{N7}
ds^2 = \Omega^2\,d\bar{s}^2\,,
\eeq
where
\beq \label{N8}
d\bar{s}^2 = -dU\,dV + \frac{4}{3}\,\Omega\,U\,dV\,dX + U^{6/5}\,e^{2\Omega X/5}\,dX^2 +U^{-2/5}\,\,e^{-2\Omega X/15}\,dY^2\,.
\eeq
The Penrose limit is obtained by letting $\Omega \to 0$ in Eq.~\eqref{N8}, namely, 
\beq \label{N9}
dS^2 = \lim_{\Omega \to 0}\,d\bar{s}^2\,.
\eeq
Hence, the Penrose limit of Harrison's TGW is
\beq \label{N10}
dS^2 = -dU\,dV + U^{6/5} \, dX^2 +U^{-2/5}\,dY^2\,,
\eeq
which represents a linearly polarized plane-wave spacetime of Petrov type \emph{N}. Indeed, it is a special case of  a class of Petrov type \emph{N} gravitational fields that represent plane waves, namely,
\begin{equation}\label{N11}
 ds^2 = - dU\,dV + U^{2\sigma_2}\,dX^2 + U^{2\sigma_3}\,dY^2\,,
\end{equation}
where  $U=T-Z$ and $V=T+Z$ are  the retarded and advanced null coordinates, respectively,  and 
\begin{equation}\label{N12}
\sigma_2+\sigma_3=\sigma_2^2+\sigma_3^2\,.
\end{equation}
Here, $\sigma_2$ and $\sigma_3$ are either both positive, or one is positive and the other is negative; moreover, if either is equal to zero or unity, this spacetime is flat. For $\sigma_2 = 3/5$ and $\sigma_3 = -1/5$, we recover the Penrose limit of Harrison's TGW spacetime. This limiting metric can be obtained in another context as well, see the paragraph containing Eq. (58) in Ref.~\cite{Bini:2018gbq}. A general discussion of metric~\eqref{N11} is contained in Section V of Ref.~\cite{Bini:2017qnd}.

Let us next return to  metric~\eqref{N8} and note that it is \emph{Ricci flat}. 
We define $\mathbb{A}$ and $\mathbb{B}$ by
\beq \label{N13}
\mathbb{A} = \frac{4}{9}\,\Omega^2\,U^{4/5}\,e^{-2\Omega X/5}\,, \qquad \mathbb{B} = \frac{2}{3}\,\Omega\,U^{-1/5}\,\,e^{-2\Omega X/5}\,
\eeq
and consider an observer in this spacetime with an orthonormal tetrad system $\lambda^{\mu}{}_{\hat \alpha}$ given in $(U, V, X, Y)$ coordinates by
\begin{eqnarray} \label{N14}
\lambda^{\mu}{}_{\hat 0} &=& (1-\mathbb{A})\,\partial_U + \partial_V - \mathbb{B}\,\partial_X\,, \nonumber \\
\lambda^{\mu}{}_{\hat 1} &=& (1+\mathbb{A})\,\partial_U - \partial_V + \mathbb{B}\,\partial_X\,, \nonumber \\
\lambda^{\mu}{}_{\hat 2} &=& U^{-3/5}\,e^{-\Omega\,X/5}\,\partial_X\,, \nonumber \\
\lambda^{\mu}{}_{\hat 3} &=& U^{1/5}\,e^{\Omega\,X/15}\,\partial_Y\,.
\end{eqnarray}
The projection of the spacetime curvature tensor on this orthonormal tetrad system
\beq \label{N15}
C_{\hat \alpha \hat  \beta \hat  \gamma \hat \delta} = C_{\mu \nu \rho \sigma}\,\lambda^{\mu}{}_{\hat \alpha}\,\lambda^{\nu}{}_{\hat \beta}\,\lambda^{\rho}{}_{\hat \gamma}\,\lambda^{\sigma}{}_{\hat \delta}\,
\eeq
can be represented by a $6\times6$ matrix $\mathcal{W} = (\mathcal{W}_{IJ})$, where the indices $I$ and $J$ range over the set $(01,02,03,23, 31,12)$. Thus we can write the measured components of the Weyl conformal curvature tensor as 
\begin{equation}
\label{N16}
\mathcal{W}=\left[
\begin{array}{cc}
\mathcal{E} & \mathcal{B}\cr
\mathcal{B} & \mathcal{-E}\cr
\end{array}
\right]\,,
\end{equation}
where $\mathcal{E}$  and $\mathcal{B}$ are symmetric and traceless $3\times3$ matrices.

The nonzero frame components of the electric part of the Weyl tensor, $\mathcal{E}_{\hat i \hat j}=C_{\hat 0\hat i\hat 0\hat j}$, are given by
\begin{eqnarray}\label {N17}
\mathcal{E}_{\hat 1\hat 1}&=&-\frac{44}{225}e^{ -\frac25 \Omega X }\Omega^2 U^{-6/5}\,, \nonumber\\
\mathcal{E}_{\hat 1 \hat 2}&=&\mathcal{E}_{\hat 2 \hat 1} =  -\frac{2}{25}\Omega U^{-8/5}e^{ -\frac15 \Omega X }-\frac{4}{75}\Omega^3 U^{-4/5}e^{ -\frac35 \Omega X }\,,\nonumber\\
\mathcal{E}_{\hat 2\hat 2}&=& \frac{6}{25 U^2}+\frac{22}{225}e^{ -\frac25 \Omega X}\Omega^2 U^{-6/5}+\frac{8}{75}\Omega^4 U^{-2/5} e^{ -\frac45 \Omega X}\,,\nonumber\\
\mathcal{E}_{\hat 3\hat 3}&=& -\mathcal{E}_{\hat 1\hat 1}-\mathcal{E}_{\hat 2 \hat 2}\,.
\end{eqnarray}
Similarly, the nonzero frame components of the magnetic  part of the Weyl tensor, $\mathcal{B}_{\hat i \hat j}=C^*_{\hat 0\hat i\hat 0\hat j}$, can be expressed as
\begin{eqnarray}\label{N18}
\mathcal{B}_{\hat 1\hat 3}&=&\mathcal{B}_{\hat 3\hat 1}= \frac{2}{25}\Omega U^{-8/5}e^{ -\frac15 \Omega X }-\frac{4}{75}\Omega^3 U^{-4/5} e^{ -\frac35 \Omega X }\,, \nonumber\\
\mathcal{B}_{\hat 2 \hat 3}&=&\mathcal{B}_{\hat 3 \hat 2}= -\frac{6}{ 25 U^2}+\frac{8}{75}\Omega^4 U^{-2/5}e^{ -\frac45 \Omega X}\,.
\end{eqnarray}

As $\Omega \to 0$, the surviving elements of the electric and magnetic parts of the Weyl tensor can be expressed as 
\begin{equation}\label{N19}
   \mathcal{E} = \mathcal{K}\,I_\oplus\,, \qquad   \mathcal{B} = -\mathcal{K}\, I_\otimes\,,\qquad \mathcal{K}(U) = \frac{6}{25\,U^2}\,,
\end{equation}
where $I_\oplus$ and $I_\otimes$ are $3\times3$ matrices defined by
\beq
\label{N20}
I_\oplus :=\left[
\begin{array}{ccc}
0&0 & 0\cr
0&1 & 0\cr
0&0 &-1\cr
\end{array}
\right]\,,\qquad
I_\otimes := \left[
\begin{array}{ccc}
0&0 & 0\cr
0&0 & 1\cr
0&1 & 0\cr
\end{array}
\right]\,,
\eeq
and represent the two (``plus" and ``cross") independent linear polarization states of gravitational radiation.

\subsection{$w$-Metric}

Next, we concentrate on the TGW spacetime given by the $w$-metric. Following Penrose, we must first write the $w$-metric~\eqref{Q1} in an appropriate form. To this end, we start with
\beq \label{P1}
ds^2 = -w^4 du\,dv + w^4\, dx^2 + w^{-2} \,dy^2\,, \qquad u := t-z\,, \qquad v :=t+z\,, \qquad w = t-z+x\,.
\eeq
As before, the null geodesic about which we take the Penrose limit is 
\beq \label{P2}
k = \partial_v\,.
\eeq
Consider now the coordinate transformation
\beq \label{P3}
(u, v, x, y) \mapsto (w, v, x, y)\,, \qquad u= w-x\,,
\eeq
which turns Eq.~\eqref{P1} into
\beq \label{P4}
ds^2 = -w^4 dw\,dv + w^4\,dv\,dx + w^4\, dx^2 + w^{-2} \,dy^2\,.
\eeq
The next step involves the coordinate transformation
\beq \label{P5}
(w, v, x, y) \mapsto (u', v', x, y)\,, \qquad u' = w^5\,, \qquad v' = \frac{1}{5}\,v\,,
\eeq
which turns the metric into
\beq \label{P6}
ds^2 = -du'\,dv' + u'^{4/5}\,(5\,dv'\,dx + dx^2) + u'^{-2/5} \,dy^2\,.
\eeq
This is the appropriate (Penrose) form for the metric. According to the Penrose procedure, we now introduce new coordinates
\beq \label{P7}
U = u'\,,\qquad \Omega^2\,V = v'\,,\qquad \Omega\,X = x\,,\qquad \Omega\,Y = y\,,
\eeq
where $\Omega$ is a positive constant. The metric then takes the form
\beq \label{P8}
ds^2 = \Omega^2\,\left[-dU\,dV + U^{4/5} \,(5\,\Omega\,dV\,dX + dX^2) +U^{-2/5}\,dY^2\right]\,.
\eeq
Let us now define the conformally related metric $d\bar{s}^2$ such that 
\beq \label{P9}
d\bar{s}^2 = \Omega^{-2}\,ds^2 = -dU\,dV + U^{4/5} \,(5\,\Omega\,dV\,dX + dX^2) +U^{-2/5}\,dY^2\,.
\eeq
In $d\bar{s}^2$, we take the limit as $\Omega \to 0$. In this way, we find
\beq \label{P10}
dS^2 = -dU\,dV + U^{4/5} \, dX^2 +U^{-2/5}\,dY^2\,,
\eeq
which also represents a linearly polarized plane-wave spacetime of Petrov type \emph{N}. Indeed, it is a special case of  Eqs.~\eqref{N11} and~\eqref{N12} with 
\beq \label{P11}
\sigma_2 = \frac{2}{5}\,, \qquad \sigma_3 = -\frac{1}{5}\,.
\end{equation}
This metric also follows from another limiting procedure involving the $w$-metric described in the paragraph containing Eq. (58) of Ref.~\cite{Bini:2018gbq}. However, an error in that paragraph should be corrected: The Harrison metric reduces to the plane wave with $(\sigma_2, \sigma_3) = (3/5, -1/5)$, while the $w$-metric reduces to $(\sigma_2, \sigma_3) = (2/5, -1/5)$.

Let us now return to Eq.~\eqref{P9} and note that this $\Omega$-dependent metric is Ricci flat as well. It proves useful to define $A'$ and $B'$ via
\beq \label{P12}
A' = \frac{25}{4}\,\Omega^2\,U^{4/5}\,, \qquad B' = \frac{5}{2}\,\Omega\,.
\eeq
Now consider an observer in this spacetime with an orthonormal tetrad system $\lambda^{\mu}{}_{\hat \alpha}$ given in $(U, V, X, Y)$ coordinates by
\begin{eqnarray} \label{P13}
\lambda^{\mu}{}_{\hat 0} &=& (1-A')\,\partial_U + \partial_V - B'\,\partial_X\,, \nonumber \\
\lambda^{\mu}{}_{\hat 1} &=& (1+A')\,\partial_U - \partial_V + B'\,\partial_X\,, \nonumber \\
\lambda^{\mu}{}_{\hat 2} &=& U^{-2/5}\,\partial_X\,, \nonumber \\
\lambda^{\mu}{}_{\hat 3} &=& U^{1/5}\,\partial_Y\,.
\end{eqnarray}
As in Eq.~\eqref{N16}, the projection of the Weyl conformal curvature tensor on this tetrad system can be expressed in terms of the gravitoelectric, $\mathcal{E}_{\hat i \hat j}=C_{\hat 0\hat i\hat 0\hat j}$,  and gravitomagnetic, $\mathcal{B}_{\hat i \hat j}=C^*_{\hat 0\hat i\hat 0\hat j}$, components of the curvature tensor as measured by the fiducial observer. We find that the nonzero gravitoelectric components can be written as
\begin{eqnarray} \label{P14}
\mathcal{E}_{\hat 1\hat 1}&=&  - 2 U^{-6/5}\Omega^2\,,\nonumber\\
\mathcal{E}_{\hat 2\hat 2}&=& \frac{6}{25 U^2} + U^{-6/5}\Omega^2+\frac{75}{8} U^{-2/5}\Omega^4\,,
\nonumber\\
\mathcal{E}_{\hat 3\hat 3}&=& -\mathcal{E}_{\hat 1\hat 1}-\mathcal{E}_{\hat 2 \hat 2}\,.
\end{eqnarray}
In a similar way, the nonzero  gravitomagnetic components are given by
\begin{eqnarray}\label{P15}
\mathcal{B}_{\hat 2\hat 3}&=&\mathcal{B}_{\hat 3\hat 2} =  -\frac{6}{25U^2} +\frac{75}{8} U^{-2/5}\Omega^4\,.
\end{eqnarray}

It seems worthwhile to use these results to illustrate the (1+1+2) decomposition of the Weyl tensor discussed in detail in Ref.~\cite{Clifton:2016mxx} and Appendix A of Ref.~\cite{TGW}. This involves, in the present case, the decomposition of our spatial, symmetric and traceless $\mathcal{E}_{\hat i \hat j}$ and $\mathcal{B}_{\hat i \hat j}$ into scalar, vector and tensor parts with respect to the unit spacelike vector $\lambda^{\mu}{}_{\hat 1}$. The tensor part is in fact a projection on the 2-dimensional screen space normal to 
$\lambda^{\mu}{}_{\hat 0}$ and $\lambda^{\mu}{}_{\hat 1}$. For the scalar parts we find
\beq \label{P16}
 \fscalarE = \mathcal{E}_{\hat 1 \hat 1} = -\frac{2\Omega^2}{U^{6/5}}\,, \qquad  \fscalarB =  \mathcal{B}_{\hat 1 \hat 1} = 0\,.
\eeq  
The vector parts vanish
\beq \label{P17}
\fvec{E}^{\hat i} = \delta^i_2\,\mathcal{E}_{\hat 2\hat 1} + \delta^i_3\,\mathcal{E}_{\hat 3\hat 1} = 0\,, \qquad \fvec{B}^{\hat i} =  \delta^i_2\,\mathcal{B}_{\hat 2\hat 1} + \delta^i_3\,\mathcal{B}_{\hat 3\hat 1} = 0\,,
\eeq  
whereas the nonzero tensor parts are given by
\begin{eqnarray} \label{P18}
\ftensor{E}_{\hat 2\hat 2} &=& -\ftensor{E}_{\hat 3\hat 3} = \frac{1}{2}\,(\mathcal{E}_{\hat 2\hat 2} - \mathcal{E}_{\hat 3\hat 3}) = \frac{6}{25 U^2} +\frac{75}{8} U^{-2/5}\Omega^4\,, \nonumber \\
\qquad  \ftensor{B}_{\hat 2 \hat3} &=& \ftensor{B}_{\hat 3\hat 2} = \mathcal{B}_{\hat 2\hat 3} =  -\frac{6}{25U^2} +\frac{75}{8} U^{-2/5}\Omega^4\,. 
\end{eqnarray}

Finally, for $\Omega \to 0$, we find exactly the same results as in the Harrison case, see Eqs.~\eqref{N19} and~\eqref{N20}. This remarkable fact can be traced back to the circumstance that for metric~\eqref{N11}, we have 
\beq \label{P19}
\mathcal{K}(U) = \frac{s_3 \,(s_3 - 1)}{U^2}\,,
\eeq
where $s_3 = -1/5$ for both Eq.~\eqref{N10} and Eq.~\eqref{P10}; in this connection, see Appendix  B of  Ref.~\cite{Bini:2017qnd}.

\section{Discussion}

We have considered the physical properties of Petrov type \emph{D} twisted gravitational wave spacetimes in this paper and investigated, in particular, their Killing vectors. 
These spacetimes admit a 4-dimensional symmetry group with a multiply transitive action on timelike hypersurfaces.
The symmetry group has a 3-dimensional subgroup having a multiply transitive action on 2-dimensional timelike surfaces which are spanned by the two principal null directions. This symmetry subgroup coincides with the Lorentz group of 2-dimensional Minkowski spacetime. This group is type $VI_0$ in the Bianchi classification.
The fourth Killing vector ($\partial_y$) is spacelike.
Furthermore, the Penrose plane-wave limit of Harrison's TGW and the $w$-metric have been explicitly determined and the corresponding transition of a TGW of Petrov type \emph{D}  with a nonplanar wave front to a plane gravitational wave of Petrov type \emph{N}  has been studied in detail.

\appendix

\section{Sectional Curvature}

Let $p$ be a point on the spacetime manifold and $u$ and $v$ be tangent vectors at $p$ that span a non-null 2-space $X$ at $p$.  Let $X := u\wedge v$; then, the sectional curvature $\Sigma_p(X)$ is defined by~\cite{BEE, HR}
\begin{equation}\label{A1}
\Sigma_p (X) = \frac{1}{2} \frac{R_{\mu \nu \rho \sigma} X^{\mu \nu}X^{\rho \sigma}}{X_{\alpha \beta} X^{\alpha \beta}}\,,\qquad X^{\mu \nu} = u^{\mu}\,v^{\nu} - u^{\nu}\,v^{\mu}\,.
\end{equation}
In the special case that the 2-space is spacelike, we can relate this definition to the $(1+1+2)$ decomposition of the Weyl tensor described near the end of Section V. Consider an observer with orthonormal tetrad 
$\lambda^{\mu}{}_{\hat \alpha}$ such that $\lambda^{\mu}{}_{\hat 0} = \theta^\mu$, $\lambda^{\mu}{}_{\hat 1} = n^\mu$, $\lambda^{\mu}{}_{\hat 2} = u^\mu$ and $\lambda^{\mu}{}_{\hat 3} = v^\mu$; that is, $X$ is  in the observer's rest space. Hence we can write $X_{\mu \nu} = \epsilon_{\mu \nu \rho \sigma}\, \theta^\rho \,n^\sigma$, where $ \epsilon_{\mu \nu \rho \sigma}$ is the Levi-Civita symbol with $ \epsilon_{\hat 0 \hat 1 \hat 2 \hat 3} := 1$ and $X_{\mu\nu} X^{\mu\nu} = 2$.  Inserting these expressions in the definition~\eqref{A1} leads to a left/right double dual of the Riemann tensor. For the Ricci-flat case under consideration in this paper, we have
\begin{equation} \label{A2}
   \Sigma_p(X) = \dual{C}^*{}_{\mu\nu\rho\sigma}\,
                      \theta^\mu\, n^\nu\, \theta^\rho\, n^\sigma
               = \dual{\quart\dual{C}}_{\mu\nu\rho\sigma}\,
                      \theta^\mu\, n^\nu\, \theta^\rho\, n^\sigma\,, 
\end{equation}
where an asterisk denotes the duality operation and we have used the equality of left and right duals of the Weyl tensor~\cite{R1}.  Furthermore,
\begin{equation}\label{A3}
   \Sigma_p(X) = \dual{\quart\dual{C}}_{\mu\nu\rho\sigma}\,
                      \theta^\mu\, n^\nu\, \theta^\rho\, n^\sigma
               = \tfrac14 \epsilon_{\mu\nu}{}^{\lambda\kappa}
                          \epsilon_{\lambda\kappa}{}^{\eta\xi}
                 C_{\eta\xi\rho\sigma}\,\theta^\mu\, n^\nu\, \theta^\rho\, n^\sigma
               = -C_{\mu\nu\rho\sigma} \,\theta^\mu\, n^\nu\, \theta^\rho\, n^\sigma\,,    
 \end{equation}
 where we have employed the identity
 \begin{equation}\label{A4}
   \epsilon_{\mu\nu}{}^{\lambda\kappa} \epsilon_{\lambda\kappa}{}^{\eta\xi}
   = -4\,\delta^\eta_{[\mu} \delta^\xi_{\nu]}\,.
\end{equation}
 Finally, 
\begin{equation}\label{A5}
   \Sigma_p(X) = - \mathcal{E}_{\hat 1 \hat 1} = -\fscalar{E}\,,
\end{equation}
which shows that the gravitoelectric scalar multiplied by $-1$ is identical to the sectional curvature of the designated spatial 2-space as defined in the $(1+1+2)$ decomposition.  

For the observer with orthonormal tetrad  $\lambda^{\mu}{}_{\hat \alpha}$, let us define 
\begin{equation}\label{A6}
  X_{\hat \alpha \hat \beta} := \lambda^{\mu}{}_{\hat \alpha}\,\lambda^{\nu}{}_{\hat \beta} - \lambda^{\nu}{}_{\hat \alpha}\,\lambda^{\mu}{}_{\hat \beta}\,,
\end{equation}
then, it is straightforward to show that 
\begin{equation}\label{A7}
\Sigma_p(X_{\hat 0 \hat i}) = - \mathcal{E}_{\hat i \hat i}\,, \qquad  \Sigma_p(X_{\hat i \hat j}) = \Sigma_p(X_{\hat j \hat i}) = R_{\hat i \hat j\hat i \hat j}\,.
\end{equation}
In Ricci-flat regions of spacetime, we find from Eq.~\eqref{N16} that $\Sigma_p(X_{\hat 0 \hat 1}) = \Sigma_p(X_{\hat 2 \hat 3})$, etc., so that these sectional curvatures are given by the measured diagonal  gravitoelectric components of the Weyl tensor multiplied by $-1$. For the Harrison TGW spacetime and the $w$-metric, the gravitoelectric components of the Weyl tensor as measured by the static observers that stay at rest in space can be obtained from the results presented in Section II of Ref.~\cite{TGW}. 

Finally, it is interesting to note that in the case of Harrison's TGW spacetime, $\Sigma_p(X_{\hat 2 \hat 3})$ calculated with respect to the orthonormal tetrad frame of the observers at rest in space, namely,
\beq \label{A8}
 \lambda_{\hat 0} = x^{-2/3}\,\partial_t\,, \qquad \lambda_{\hat 1}= x^{-2/3}\,\partial_z\,, \qquad \lambda_{\hat 2} = u^{-3/5}\, \partial_x\,,\qquad 
\lambda_{\hat 3}= x^{1/3} u^{1/5}\, \partial_y\,,
\eeq
turns out to be the Gaussian curvature $K_G$ of the wave front~\cite{Bini:2018gbq, TGW}; that is, 
\begin{equation} \label{A9}
\Sigma_p(X_{\hat 2 \hat 3}) = K_G = -\frac{4}{9}\,\frac{1}{u_0^{6/5}\,x^2}\,,
\end{equation}
where $u = u_0$ is the wave front. Similarly, for the $w$-metric the corresponding result is~\cite{Bini:2018gbq, TGW}
\begin{equation} \label{A10}
\Sigma_p(X_{\hat 2 \hat 3}) = K_G = -\frac{4}{(u_0 + x)^6}\,.
\end{equation}

\section*{Acknowledgments}

D.B. thanks the Italian INFN (Naples) for partial support.

\end{document}